\documentclass[aps,prl,amsmath,amssymb,reprint,superscriptaddress,]{revtex4-2}
\usepackage{charter,graphicx,verbatim,threeparttable,float,amssymb }
\usepackage[version=4]{mhchem}

\begin{document}

\title[CVSStripes]{Incommensurate charge-stripe correlations in the kagome superconductor CsV$_3$Sb$_{5-x}$Sn$_x$}

\author{Linus Kautzsch}
\affiliation{Materials Department, University of California Santa Barbara, California 93106 United States}
\author{Yuzki M. Oey}
\affiliation{Materials Department, University of California Santa Barbara, California 93106 United States}
\author{Hong Li}
\affiliation{Department of Physics, Boston College, Chestnut Hill, MA 02467, USA}
\author{Zheng Ren}
\affiliation{Department of Physics, Boston College, Chestnut Hill, MA 02467, USA}
\author{Brenden R. Ortiz}
\affiliation{Materials Department, University of California Santa Barbara, California 93106 United States}
\author{Ram Seshadri}
\affiliation{Materials Department, University of California Santa Barbara, California 93106 United States}
\author{Jacob Ruff}
\affiliation{CHESS, Cornell University, Ithaca, New York 14853, USA}
\author{Ziqiang Wang}
\affiliation{Department of Physics, Boston College, Chestnut Hill, MA 02467, USA}
\author{Ilija Zeljkovic}
\affiliation{Department of Physics, Boston College, Chestnut Hill, MA 02467, USA}
\author{Stephen D. Wilson}
\affiliation{Materials Department, University of California Santa Barbara, California 93106 United States}
\email{stephendwilson@ucsb.edu}
\date{\today}

\begin{abstract}
We track the evolution of charge correlations in the kagome superconductor CsV$_3$Sb$_5$ as its parent, long-ranged charge density order is destabilized. Upon hole-doping doping, interlayer charge correlations rapidly become short-ranged and their periodicity is reduced by half along the interlayer direction. Beyond the peak of the first superconducting dome, the parent charge density wave state vanishes and incommensurate, quasi-1D charge correlations are stabilized in its place. These competing, unidirectional charge correlations demonstrate an inherent electronic rotational symmetry breaking in CsV$_3$Sb$_5$, independent of the parent charge density wave state and reveal a complex landscape of charge correlations across the electronic phase diagram of this class of kagome superconductors. Our data suggest an inherent 2$k_f$ charge instability and the phenomenology of competing charge instabilities is reminiscent of what has been noted across several classes of unconventional superconductors.
\end{abstract}

\maketitle

Charge correlations and the nature of charge density wave (CDW) order within the new class of $A$V$_3$Sb$_5$ ($A$=K, Rb, Cs) kagome superconductors \cite{ortiz2019new,ortiz2020cs,ortiz2021superconductivity,yin2021superconductivity} are hypothesized to play a crucial role in the anomalous properties of these compounds.  Hints of pair density wave superconductivity \cite{chen2021roton,ge2022discovery} as well as signatures of orbital magnetism \cite{xu2022universal,mielke2022time,yu2021evidence,guo2022field} in these compounds are all born out of a central CDW state \cite{jiang2021unconventional,zhao2021cascade,shumiya2021intrinsic}. The CDW order parameter itself is theorized to host both primary, real and secondary, imaginary components \cite{park2021electronic}, each of which thought to play a role in the anomalous properties observed in $A$V$_3$Sb$_5$ compounds. 

The real component of the CDW state in $A$V$_3$Sb$_5$ compounds manifests primarily as a $2\times2$ reconstruction within the kagome plane driven via a 3\textbf{q} distortion into either star-of-David (SoD) or (its inverse) tri-hexagonal (TrH) patterns of order \cite{tan2021charge}. In-plane 3\textbf{q} distortions are further correlated between kagome layers \cite{liang2021three,ortiz2021fermi,li2021observation,jiang2021unconventional}, either through correlated phase shifts of the same distortion type between neighboring layers, via alternation between distortion mode types, or a combination of both \cite{christensen2021theory}. 

The parent CDW state of CsV$_3$Sb$_5$ forms as a $2\times2\times4$ supercell whose average structure is comprised of a modulation between SoD and TrH distortion modes along the interlayer $c$-axis below $T_{CDW}=94$ K \cite{ortiz2021fermi,kang2022microscopic,hu2022coexistence}.  Upon cooling, however, the CDW state also shows hints of a staged behavior, suggestive of another coexisting or competing CDW instability.  Scanning tunneling microscopy (STM) measurements resolve commensurate, quasi-1D charge stripes that form near $T\approx 60$ K and coexist with the $2\times2$ in-plane CDW order \cite{zhao2021cascade}. Transient reflectivity \cite{ratcliff2021coherent} and Raman measurements \cite{wu2022charge} also resolve a shift/new modes in the lattice dynamics near this same energy scale.  Sb NQR and V NMR measurements further resolve a chemical shift below 60 K \cite{luo2022possible}, demonstrating a structural response to a modified CDW order parameter---a response potentially driven by a competing CDW instability.  

Further supporting the notion of a competing, electronic instability is the rapid suppression of the parent CDW order in CsV$_3$Sb$_5$ under moderate pressure \cite{chen2021double,yu2021unusual} or via small levels of hole-substitution \cite{oey2022fermi}. By substituting $\approx 6$ $\%$ holes per formula unit, the CDW state seemingly vanishes in thermodynamic measurements, while superconductivity undergoes a nonmonotonic response and generates two superconducting domes.  The evolution of charge correlations as the parent, three-dimensional CDW order is suppressed via hole substitution therefore stands to provide important insights into underlying or competing electronic instabilities and their roles in the unconventional coupling between CDW order and superconductivity reported in CsV$_3$Sb$_5$.

\begin{figure}
	\includegraphics[scale=0.7]{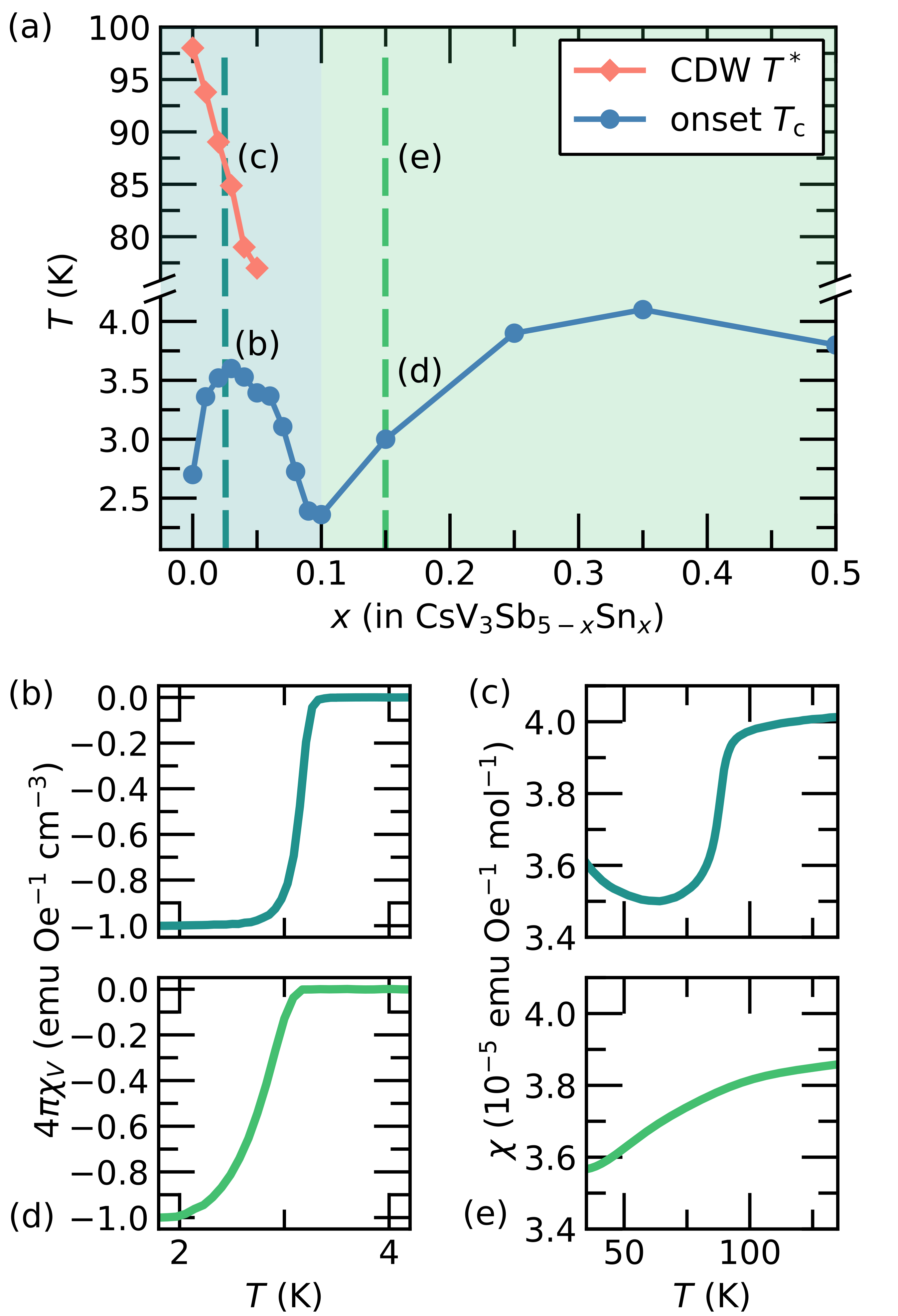}
	\caption{(a) Electronic phase diagram of Sn-doped CsV$_3$Sb$_5$ showing the evolution of both CDW and SC order with hole-doping.  Data are reproduced from Ref. \cite{oey2022fermi}.  Panels (b) and (c) show susceptibility data characterizing the superconducting and CDW states of the $x=0.025$ composition in the first SC ``dome" and panels (d) and (e) show susceptibility data characterizing the superconducting and CDW states of the $x=0.15$ composition in the second SC ``dome".}
	\label{fig:1structure}
\end{figure}

Here we track the evolution of charge correlations in CsV$_3$Sb$_{5-x}$Sn$_{x}$ as holes are introduced via Sn-substitution and the parent $2\times2\times4$ CDW state is suppressed.  X-ray diffraction data resolve that very light Sn-substition ($x=0.025$) suppresses interlayer correlations, and the CDW immediately becomes $2\times2\times2$ with short-range correlations along the c-axis.  Increased hole-doping reveals the continued shortening of interlayer correlations and the eventual suppression of in-plane $2\times2$ CDW order; however, at concentrations where the parent CDW state is suppressed ($x=0.15$), incommensurate quasi-1D charge correlations appear.  Parallel STM measurements at this same composition also observe the persistence of low-temperature quasi-1D charge stripes in the absence of rotational symmetry breaking $2\times2$ CDW order \cite{li2022rotation}. Our data demonstrate an underlying electronic rotational symmetry breaking distinct from the broken rotational symmetry arising from the commensurate $2\times2\times4$ CDW state in the parent material, directly unveiling a complex landscape of charge correlations whose competition is reflected in the superconducting transition temperature.

 CsV$_3$Sb$_{5-x}$Sn$_{x}$ crystals with $x$ = 0.025 and $x$ = 0.15 were made with a flux of Cs$_{20}$V$_{15}$Sb$_{90}$Sn$_{30}$ and Cs$_{20}$V$_{15}$Sb$_{106}$Sn$_{34}$ respectively. Fluxes were ball-milled for 60 mins and then packed into alumina crucibles, and sealed under inert atmosphere within stainless steel tubes. Tubes were heated to 1000 $^\circ$C and kept at 1000 $^\circ$C for 12 hours and then cooled quickly to 900 $^\circ$C, and then slowly cooled (2 $^\circ$C / hour) to 500 $^\circ$C. Temperature-dependent synchrotron x-ray diffraction were collected at the ID4B (QM2) beamline, CHESS. In ID4B measurements, temperature was controlled by a stream of cold helium gas flowing across the single-crystal sample. An incident x-ray of energy 26 keV ($\lambda=0.6749 $ \AA) was selected using a double-bounce diamond monochromator. Bragg reflections were collected in transmission mode, and the sample was rotated with full 360$^\circ$ patterns, sliced into 0.1$^\circ$ frames.  STM data were acquired using a Unisoku USM1300 STM at approximately 4.5 K. Spectroscopic measurements were made using a standard lock-in technique with 915 Hz frequency and bias excitation. STM tips were custom-made chemically-etched tungsten tips, annealed in UHV to bright orange color prior to the measurements.

To understand the evolution of charge correlations across the electronic phase diagram of CsV$_3$Sb$_{5-x}$Sn$_{x}$, two Sn concentrations were chosen as shown in Fig. 1 (a).  The first $x=0.025$ concentration possesses both a superconducting state with an enhanced $T_c$ and a clearly observable CDW transition as shown in Figs. 1 (b) and (c).  The second $x=0.15$ concentration retains a SC phase transition but the thermodynamic signature of $2\times2$ CDW order in magnetization data has vanished as shown in Figs. 1 (d) and (e).  

\begin{figure}
	\includegraphics[scale=0.55]{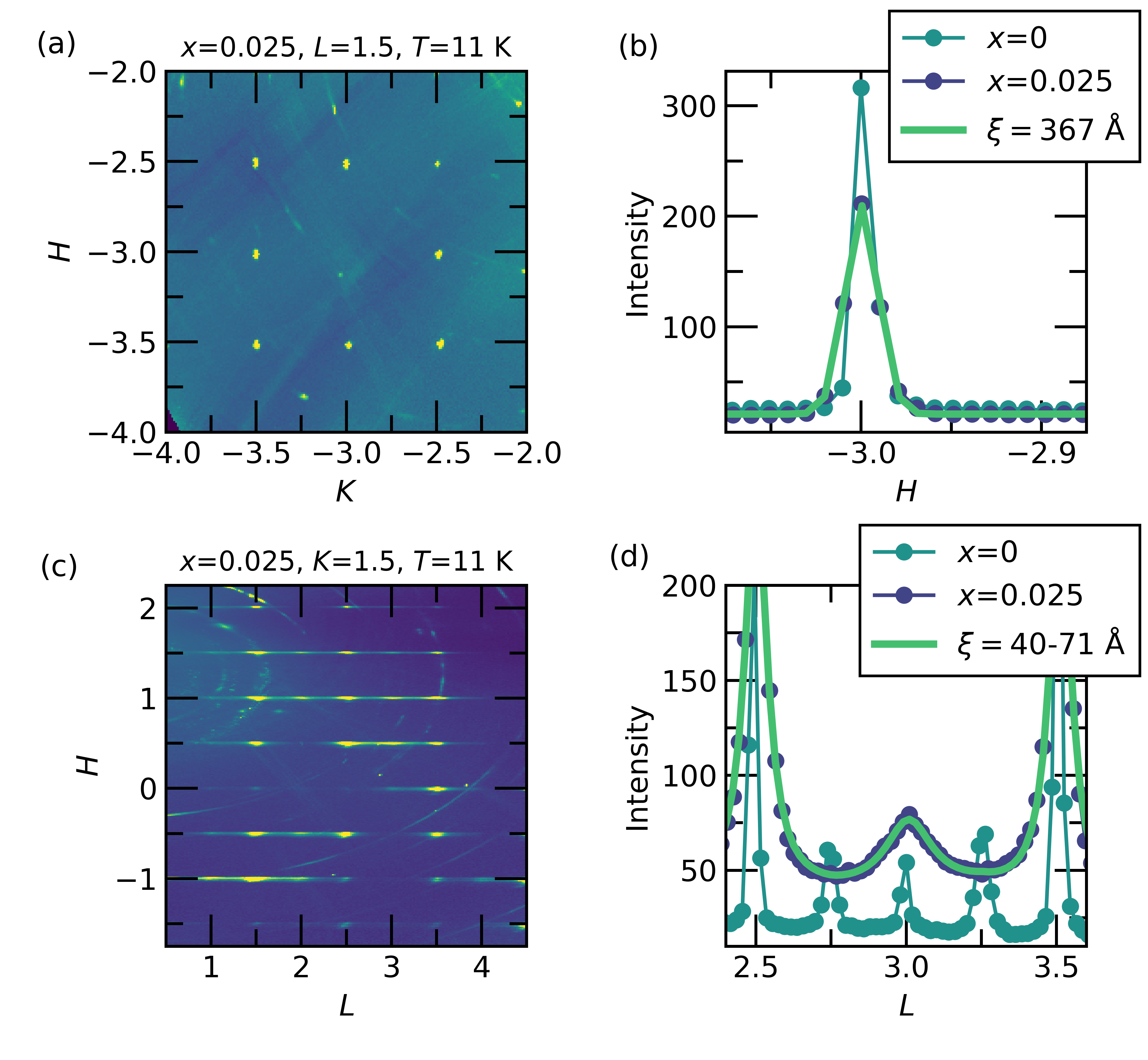}
	\caption{(a) Map of x-ray scattering intensities in the ($H$, $K$, 1.5)-plane for the $x=0.025$ sample at $T=11$ K.  (b) One dimensional $H$-cuts through the (-3, -2.5, 1.5) position for both $x=0$ and $x=0.025$.  Solid lines are Gaussian fits to the data.  (c) Map of x-ray scattering intensities in the ($H$, 1.5, $L$) plane for the $x=0.025$ sample.  (d)  One-dimensional $L$-cuts along $H$=1 for both the $x=0$ and $x=0.025$ samples.  Solid lines are pseudo-Voigt fits for the $x=0.025$ sample with the Gaussian component fixed to the instrument's resolution.}
	\label{fig:2SCCDWdata}
\end{figure}

Looking first at the $x=0.025$ crystal, maps of x-ray diffraction data were collected with representative data plotted in Figs. 2 (a) and (b). Fig. 2 (a) plots scattering within the ($H$, $K$, 1.5)-plane.  Reflections centered at ($H$, $K$)=(0.5, 0.5)-type positions indicate that the $2\times2$ in-plane CDW order present in the undoped CsV$_3$Sb$_5$ remains in the $x=0.025$ compound.  Interlayer correlations are, however, altered. Fig. 2 (c) plots scattering within the ($H$, 1.5, $L$)-plane, showing that $c$-axis correlations shift to substantially shorter-range and center at the $L=0.5$ position.  This marks a switch into a quasi-2D $2\times2\times2$ CDW state from the long-range $2\times2\times4$ order of the undoped material and a transition into a CDW state whose \textbf{q} vectors match those of undoped (K,Rb)V$_3$Sb$_5$ \cite{jiang2021unconventional}.

The in-plane correlation lengths associated with CDW peaks in the $x=0.025$ sample are slightly reduced, shortening from resolution-limited in the undoped material to $\xi_{H}=367\pm6$ \AA. Interplane correlation lengths shorten dramatically, moving from resolution-limited again in the undoped material to $\xi_{L}=70\pm2$ \AA.  Weak reflections also persist at $L=$ integer positions with shorter correlation lengths $\xi_{L}=40\pm2$ \AA.  The presence of these integer $L$ reflections likely reflects that interlayer correlations continue to be heavily impacted by the nearly degenerate array of states along the $U$-line of the Brillouin zone \cite{wu2022charge,xiao2022coexistence,stahl2022temperature}, and the difference in correlation lengths between $L=0.5$ and $L=0$ type positions reflects two distinct patterns of order in the doped sample forming as it transitions to being truly quasi-2D.

At these small doping levels, while still in the CDW state, the immediate disappearance of $L$=0.25 type peaks upon small amounts of impurity/carrier substitution suggests a rapid crossover in the character of CDW order of CsV$_3$Sb$_5$ and a doping-induced switch from a state with modulating star-of-David and tri-Hexagonal order into one with phase shifted planes of a single distortion type, similar to (K,Rb)V$_3$Sb$_5$ \cite{kang2022microscopic}.  This crossover into another CDW phase at light doping may drive the formation of the first SC dome in the phase diagram of CsV$_3$Sb$_{5-x}$Sn$_x$; however a quantitative refinement of the intermediate $2\times2\times2$ CDW state will be required to further understand the mechanism.

Now examining charge correlations outside of the CDW phase boundary in the electronic phase diagram, x-ray scattering data for the $x=0.15$ sample are plotted in Fig. 3.  Panels (a) and (b) show a representative schematic of the scattering about the zone center and the corresponding data ($H$, $K$, -0.5)-plane.  Data collected at $\frac{1}{2}$-integer $L$ values indicate a superposition of three quasi-1D patterns of charge scattering, rotated $60^{\circ}$ from one another in reciprocal space.  This superposition of quasi-1D correlations can be understood in a model of charge correlations forming along one unique crystal axis (i.e. $H$ or $K$), reducing the six-fold rotational symmetry of the lattice to two-fold, and forming three one-dimensional domains rotated by $120^{\circ}$ in real space.  This rotational symmetry breaking vanishes upon warming as shown in Fig. 3 (d), consistent with domain formation upon cooling. 

\begin{figure}
	\includegraphics[scale=0.6]{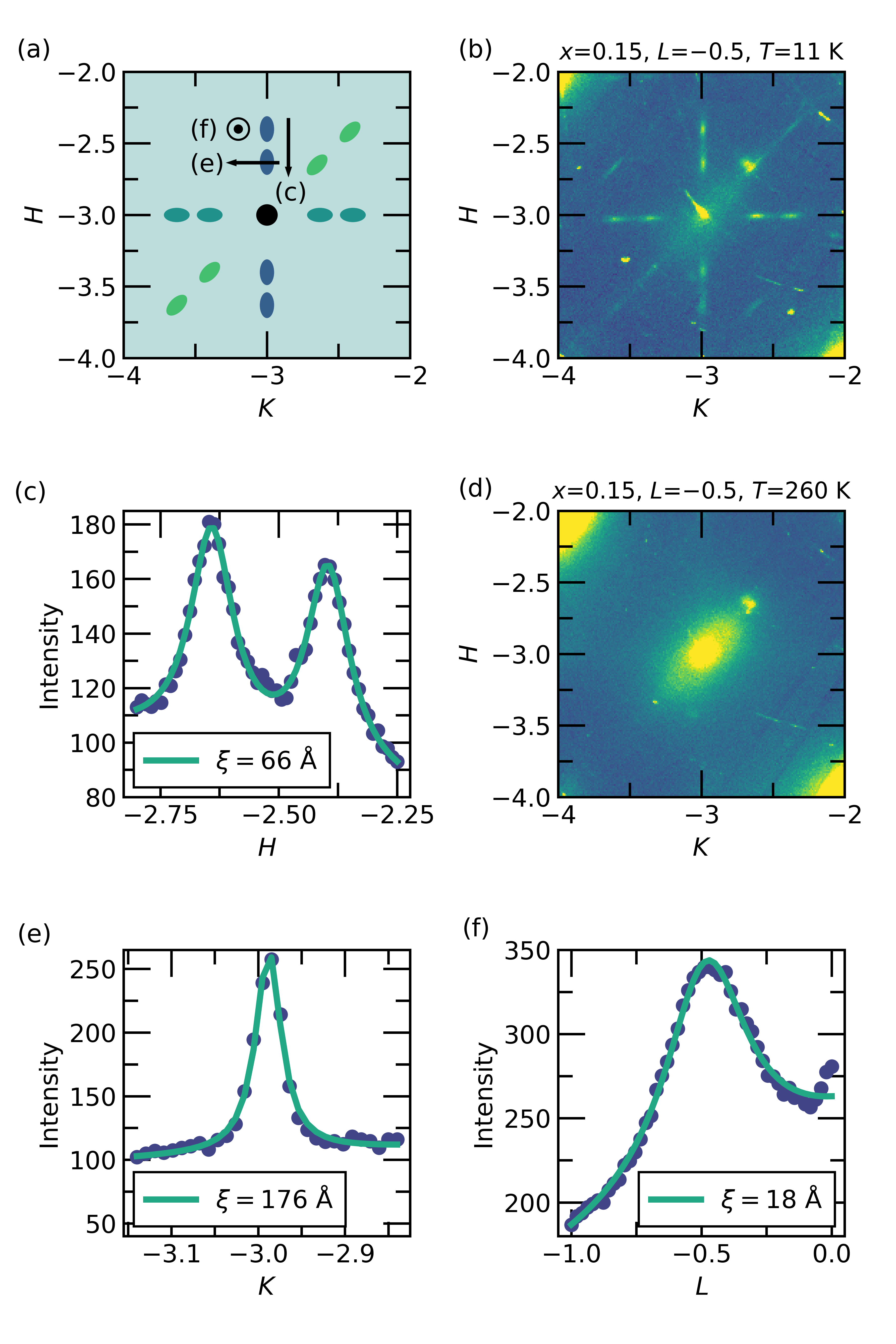}
	\caption{(a) Schematic of x-ray scattering in the ($H$, $K$)-plane about a representative zone center for the $x=0.15$ sample.  Scattering from three domains is illustrated and cut directions for corresponding panels are labeled. (b) Map of x-ray scattering intensities for $x=0.15$ at $T=11$ K plotted about ($H$, $K$, -0.5) (c) One dimensional cut along $H$ as illustrated in panel (a), (d) Map of x-ray scattering intensities for $x=0.15$ at $T=300$ K (e-f) One dimensional cuts along $K$ and $L$ as illustrated in panel (a).  Solid lines are the results of pseudoVoigt fits to the peak lineshapes with the Gaussian component constrained to the instrument's resolution.}
	\label{fig:3phasediagram}
\end{figure}

Looking at scattering from a single quasi-1D domain, the charge correlations form an incommensurate state with \textbf{q$_{inc}$}=0.37 along a preferred in-plane axis.  This is illustrated via a representative cut along $H$ plotted in Fig. 3 (c).  Within the ($H$, $K$)-plane, correlations along \textbf{q$_{inc}$} are short-ranged with $\xi_{H}=66\pm2$ \AA ~ and are substantially longer-ranged orthogonal to the direction of modulation with $\xi_{K}=176\pm7$ \AA ~ (Fig. 2(e)).  As shown in Fig. 3 (f), the peak of these quasi-1D correlations is centered at the $L=-0.5$ position with a short-correlation length of $\xi_{L}=18\pm1$ \AA, reflecting an anti-phase modulation between neighboring kagome layers correlated only between neighboring V-planes.  

\begin{figure*}
	\includegraphics[scale=0.65]{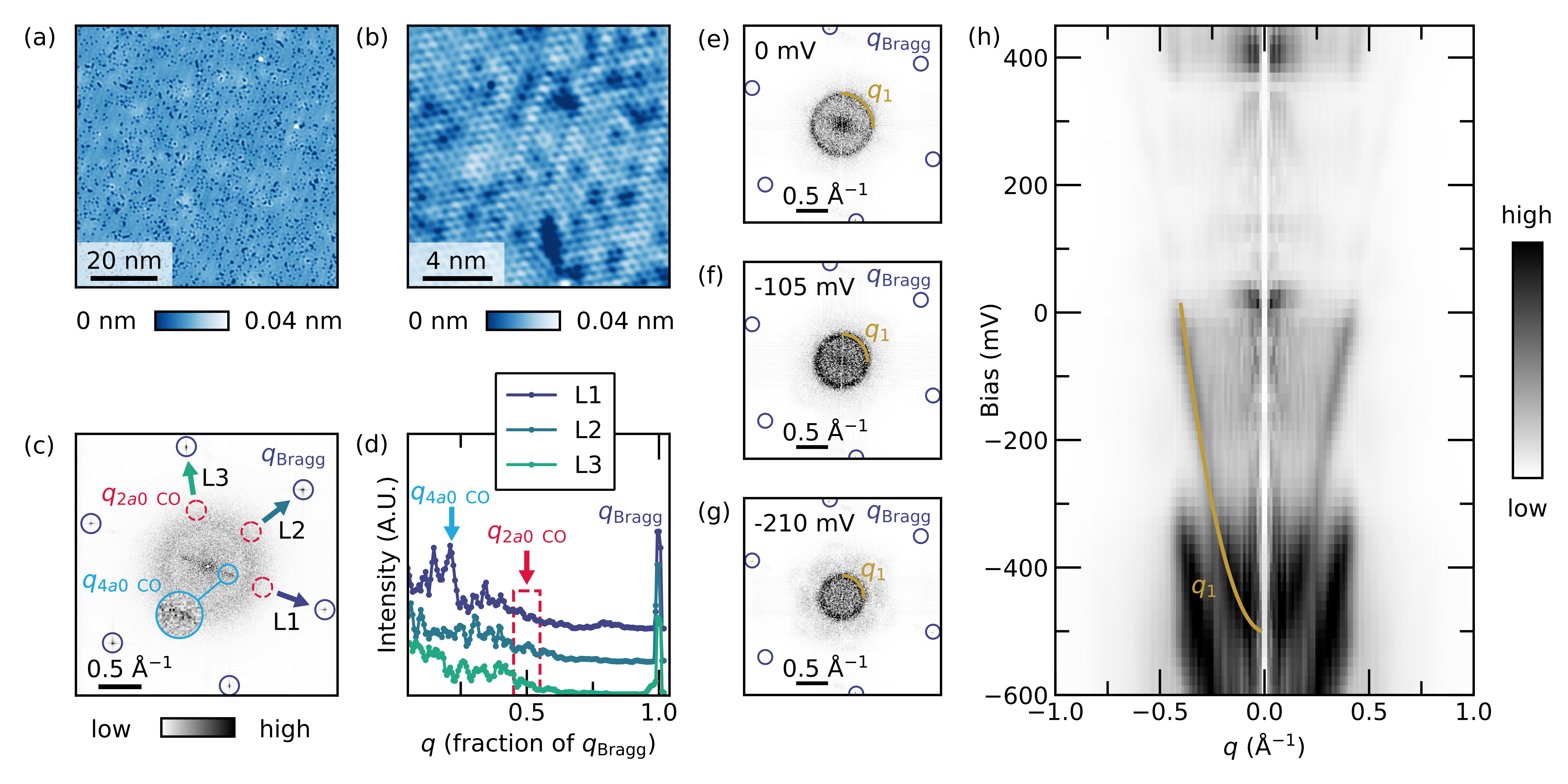}
	\caption{(a) and (b) show STM topography images of CsV$_3$Sb$_5$, (c) Fourier transform of the STM topography showing the presence of $4a_0$ order and the absence of $2a_0$ correlations, (d) One dimensional line cuts through the Fourier map in panel (c), (e-g) Quasiparticle interference spectra collected at 0 mV, -105 mV, and -210 mV biases respectively.  The circular scattering from \textbf{$q_1$} due to the Sb $p_z$ states is marked. (h) The dispersion of the QPI pattern showing the bottom of the Sb $p_z$ band has risen to $\approx -500$ meV.}
	\label{fig:4dft}
\end{figure*}

The appearance of incommensurate quasi-1D charge correlations following the collapse of 3D CDW order is suggestive of an underlying incommensurate and unidirectional instability whose correlations compete with the parent $2\times 2$ in-plane CDW order.  Quasi-1D fluctuations were previously observed in STM measurements to onset well below $T_{CDW}$ in undoped CsV$_3$Sb$_5$ \cite{zhao2021cascade}, consistent with this notion of a nearby quasi-1D state. These fluctuations in the $x=0$ system are locally pinned at the surface into a nearly commensurate surface phase and coherent, quasi-1D band features appear in the differential conductance d$I$/d$V$ maps \cite{li2022emergence}, reflective of a strong coupling between these fluctuations and the electronic structure.  

To further investigate the local evolution of charge correlations, we perform STM measurements on the $x=0.15$ sample at 4.5 K.  Figs. 4 (a) and (b) show STM topographs of the Sb surface over different fields of view where dark hexagonal defects correspond to individual Sn dopants.  Counting these defects is consistent with the expected Sn concertation of 0.15 Sn atoms per formula unit. One-dimensional, stripe-like features are apparent in the STM topograph (Figs. 4 (a) and (b)), which can be more easily quantified via the Fourier transform plotted in Fig. 4 (c). In this Fourier map, quasi-1D correlations are observed along one of the atomic Bragg peak directions, similarly to the previously identified $4a_0$ charge stripes in the $x=0$ sample \cite{zhao2021cascade}. The superlattice peaks at the $2\times2$ (or $2a_0$) CDW positions are notably absent.  This is further demonstrated via the line cuts through the Fourier map along the three lattice directions, where no scattering peaks can be observed at 0.5Q$_{Bragg}$ reciprocal space positions (Fig. 4 (d)). 

To gain insight into the electronic band structure of the system, QPI imaging is plotted in Fig. 4.  Fourier transforms of STM d$I$/d$V$ maps in Figs. 4 (e)-(g) show the electron scattering and interference pattern as a function of increasing STM bias (binding energy). The dominant dispersive scattering wave vector is the nearly isotropic central circle (labeled \textbf{$q_1$}), which arises from scattering within the Sb $p_z$ band that crosses through $E_f$.  Hole-doping is predicted to be orbitally-selective and should preferentially dope this band \cite{labollita2021tuning,oey2022fermi}, pushing the bottom of the band closer to $E_f$. Figure 4 (h) shows the resulting dispersion of \textbf{$q_1$}, where it can be seen that the bottom of the Sb $p_z$ band has been pushed up from below -600 meV in the $x=0$ parent system \cite{zhao2021cascade} to $\approx -500$ meV in the $x=0.15$ sample. This is consistent with DFT expectations of hole-doping achieved via the replacement of in-plane Sb atoms with Sn. 

The persistence of quasi-1D charge correlations on the surface of the $x=0.15$ sample in the absence of the $2\times2$ CDW state suggests that the fluctuations driving this surface order are robust to hole-doping and are likely linked to the incommensurate quasi-1D correlations that stabilize in x-ray scattering measurements once long-range CDW order is suppressed.  The trade-off between the two types of charge correlations in bulk x-ray measurements suggests that an underlying nematicity is operative across the electronic phase diagram of CsV$_3$Sb$_5$ (i.e. beyond where long-range CDW order vanishes). Our experiments establish $A$V$_3$Sb$_5$ as a promising platform for the studies of charge-stripe physics by scattering experiments and draw comparisons with the extensively studied 4$a_0$ charge ordering in cuprates \cite{doi:10.1146/annurev-conmatphys-031115-011401}. For example, the sizable doping dependence of charge ordering in Bi-based cuprates \cite{da2014ubiquitous} appears qualitatively similar to what we have observed here in CsV$_3$Sb$_5$ with Sn doping. Given the suppression of charge ordering in cuprates in the overdoped regime, it will be of interest to explore the fate of 1D charge correlations in CsV$_3$Sb$_5$ at an even higher doping levels, as samples with higher Sn composition are developed in the future.

In summary, our results demonstrate a complex landscape of charge correlations in the hole-doped kagome superconductor CsV$_3$Sb$_{5-x}$Sn$_x$.  Light hole-doping drives a transition into a neighboring CDW state with a $2\times2\times2$ supercell and short-range interlayer correlations.   Doping further holes results in the suppression of the 3\textbf{q} CDW state and the striking stabilization of quasi-1D, incommensurate charge correlations. These emergent quasi-1D correlations demonstrate an underlying electronic rotational symmetry breaking present across the phase diagram of this system and are suggestive of a 2$k_f$ instability at the Fermi surface.  Our results provide important experimental insights into neighboring/hidden charge correlations in the new class of $A$V$_3$Sb$_5$ superconductors and provide crucial input for modeling the unconventional interplay between charge density wave order and the low-temperature superconducting ground state.

\section{acknowledgments}
This work was supported by the National Science Foundation (NSF) through Enabling Quantum Leap: Convergent
Accelerated Discovery Foundries for Quantum Materials Science, Engineering and Information (Q-AMASE-i): Quantum Foundry at UC Santa Barbara (DMR-1906325). I.Z. gratefully acknowledges the support from the National Science Foundation grant NSF-DMR-1654041.  Z.W. is supported by U.S. Department of Energy, Basic Energy Sciences Grant No. DE-FG02-99ER45747 and the Cottrell SEED Award No. 27856 from Research Corporation for Science Advancement.  The research reported here made use of shared facilities of the NSF Materials Research Science and Engineering Center at UC Santa Barbara DMR-1720256, a member of the Materials Research Facilities Network (www.mrfn.org). This work is based upon research conducted at the Center for High Energy X-ray Sciences (CHEXS) which is supported by the National Science Foundation under award DMR-1829070.  Any opinions, findings, and conclusions or recommendations expressed in this material are those of the authors and do not necessarily reflect the views of the National Science Foundation. 

\section*{References}
\bibliography{SnDopedBib}

\end{document}